**Dirac-point photocurrents due to photothermoelectric effect in non-uniform graphene devices**

Michael S. Fuhrer[1,2*] and Nikhil V. Medhekar[1,3]

[1]*ARC Centre of Excellence in Future Low Energy Electronic Technologies, Monash University, Monash 3800 Victoria, Australia*
[2]*School of Physics and Astronomy, Monash University, Monash 3800 Victoria, Australia*
[2]*Department of Materials Science and Engineering, Monash University, Monash 3800 Victoria, Australia*
*michael.fuhrer@monash.edu

Q. Ma et al.[1] recently reported a strong photocurrent associated with charge neutrality in graphene devices with non-uniform geometries, which they interpreted as an intrinsic photoresponse enhanced by the momentum non-relaxing nature of electron-electron collisions at charge neutrality. Here we argue that gradients in charge carrier density give rise to a photothermoelectric effect (PTE) which is strongly peaked around charge neutrality, i.e. at *p-n* junctions, and such *p-n* junctions naturally arise at the edges of graphene devices due to fringing capacitance. Using known parameters, the PTE effect in the presence of charge density gradients predicts the sign, spatial distribution, gate voltage dependence, and temperature dependence of the photoresponse in non-uniform graphene devices, including predicting the observed sign change of the signal away from charge neutrality, and the non-monotonic temperature dependence, neither of which is explained by the intrinsic photocurrents in graphene. We propose future experiments which may disentangle the contributions of PTE and intrinsic photocurrent in graphene devices.

Q. Ma et al.[1] show that the PTE is zero in the case of uniform carrier density *n* in graphene. However, if *n* is non-uniform, a local PTE arises, proportional to the gradient in the local thermopower: $\nabla S = (\frac{dS}{dn})\nabla n$. One known source of such non-uniformity is the variation in the local backgate capacitance to the graphene. The effect of geometry on capacitance is significant; for example, the capacitance per area for a 1 micron strip of graphene on 300 nm $SiO_2$ is ~40% larger than for a 10 micron strip of graphene[3] largely due to fringing capacitance at the strip edges.

Fig. 1 models the non-uniform carrier density in a graphene device of geometry similar to Fig. 2 in [1]. As-fabricated graphene devices often have a uniform background carrier density -$n_D$ without application of a gate voltage, and require a finite backgate voltage $V_{BG} = V_D = n_D e/c_g$, where $c_g$ is the average capacitance per unit area and *e* the elementary charge, to tune to charge neutrality. Fig. 1 shows that at $V_{BG} = V_D$, the charge density is non-uniform in graphene, and a *p-n* junction exists near and roughly parallel to the graphene edge. Such a *p-n* junction will give rise to a local photocurrent due to PTE when illuminated by light, and this in turn drives a global current, independent of distance to the collecting electrodes, according to the Shockley-Ramo framework developed in [2]. Since the *p-n* junction closely follows the graphene edges (Fig. 1) the spatial dependence of the photocurrent is indistinguishable from the model in [1]. Other sources of non-uniform doping, such as chemical termination of edges, treatment of the oxide near the edge during plasma etching, or self-doping from edge states could also give rise to *p-n* junctions at graphene edges independent of the gating effect, but will not be discussed here.

We now discuss the sign of the photocurrent. For the device in Fig. 2 of [1], $R(V_{BG})$ is shown, indicating $V_D$ is positive, approximately +15 V (Fig. 2e). At $V_{BG} = V_D$ the device is relatively *n*-doped at edges and *p*-doped in the interior. Thus illumination will drive a (positive) photocurrent towards the *p*-type region, which is in agreement with the sign of the photocurrent in Fig. 2 of [1]. Note that the sign of the intrinsic photocurrent proposed in [1] results from slight electron-hole asymmetry in the bandstructure, and does not depend on the sign of $V_D$. Interestingly the sign of the Dirac-point photocurrent in Fig. 1 of [1] is opposite to that in Fig. 2 of [1](i.e. on illumination positive current flows from wide to narrow graphene in Fig. 1, but from narrow to wide graphene in Fig. 2) which is unexplained by the intrinsic photocurrent model. The sign of $V_D$ is not given for the device in Fig. 1 of [1].

We now model the gate voltage dependence of the PTE for the device in Fig. 2 in [1]. We assume the conductivity σ ~|n| and Fermi energy $E_F \sim n^{1/2}$, where $n$ is carrier density, and then convolute σ(n) and $E_F(n)$ with a Gaussian of width $n^* = 1.8 \times 10^{11}$ cm$^{-2}$ to model the effect of puddling and reproduce the width of the resistivity peak. We assume $n = 7.2 \times 10^{10}$ cm$^{-2}$($V_{BG} - V_D$), and $V_D = +15$ V. We calculate $S$ based on the Mott relation $S = -\frac{\pi^2 k_B^2 T}{3e} \frac{1}{\sigma} \frac{d\sigma}{dE_F}$ and assume the PTE signal is proportional to $V_{BG}(dS/dV_{BG})$; the proportionality to $V_{BG}$ reflects that the density gradient itself is established by $V_{BG}$ through the non-uniform capacitance.

Fig. 2 shows the $R(V_g)$ and PTE signal from our model, as well as the measured $R(V_g)$ and photocurrent from [1]. The agreement between experiment and model is extraordinary, especially given that there are no free parameters in the model beyond those used to match $R(V_g)$ (width $n^*$ and Dirac point shift $V_D$). We see that the modelled photocurrent is peaked strongly around charge neutrality (one of the most striking features of the experiment), goes to zero around the half-width of the resistivity peak, and reverses sign at values of $V_{BG}$ farther from the charge neutrality point. The sign reversal is a unique feature of the PTE arising from the non-monotonic $S(V_{BG})$; this sign reversal was critical in the first conclusive demonstration of the PTE in graphene[4]. The sign reversal is seen in the experiment (this can also be seen as reverse-colour lobes around each central lobe in Fig. 2d of [1] not shown). The maximum signal in the sign-reversed region is much smaller than in the central peak, but larger at larger $V_{BG}$ than smaller $V_{BG}$. The PTE model reproduces all of the features of the experiment, including the sign change. Notably, the intrinsic photoresponse model cannot explain this sign change.

The photoresponse in [1] was reported to be non-monotonic in temperature, peaking at ~100 K. A strikingly similar non-monotonic temperature dependence has been measured for the photocurrent in intentionally fabricated graphene $p$-$n$ junctions[5], resulting from the competition of hot electron relaxation processes, dominated by intrinsic acoustic phonon emission at low temperature and extrinsic disorder-mediated scattering at high temperature[5]. The crossover is dependent on disorder and carrier density, but the measured temperature dependence for the PTE of a graphene $p$-$n$ junction at low carrier density[5] is very similar to that observed in [1] though peaks at a slightly lower temperature ~60 K.

We conclude that the known PTE effect can naturally explain many of the observations in [1] in graphene with non-uniform geometries, with no need to invoke an additional photocurrent mechanism. However, in their supplemental materials, Q. Ma et al. also show photocurrent experiments on other graphene devices, some with more regular geometries. One device exhibits a Dirac-point photocurrent at $V_{BG} = 0$, which is not explained by our model of a gate-induced edge $p$-$n$ junction. Some devices, such as that in S3.4(1), include a $p$-$n$ junction and should (and do) exhibit the density-gradient PTE discussed here. Others do not have any intentional $p$-$n$ junctions, or exhibit photocurrents away from the $p$-$n$ junctions. In these latter devices, some inhomogeneity is needed to explain the photocurrent by any mechanism, whether PTE or intrinsic. More work is needed to understand the nature of this inhomogeneity (whether in carrier density or from some other source) and how it gives rise to photocurrent.

Additional experiments could be designed to discriminate more convincingly between PTE and intrinsic photocurrent mechanisms, for example by eliminating the effects of stray geometric capacitance (using very thin back gate dielectrics or adding an additional ground plane adjacent to graphene). Locally gating the graphene edge with an adjacent electrode could be particularly enlightening, as the PTE effect is expected to depend on the sign of the edge doping, while the sign of the intrinsic photocurrent is expected to be determined only by electron-hole asymmetry in the bandstructure.

**Acknowledgements**

The authors acknowledge support of the ARC Centre of Excellence FLEET (CE170100039).

**Data Availability**

The datasets generated during and/or analysed during the current study are available from the corresponding author on reasonable request.

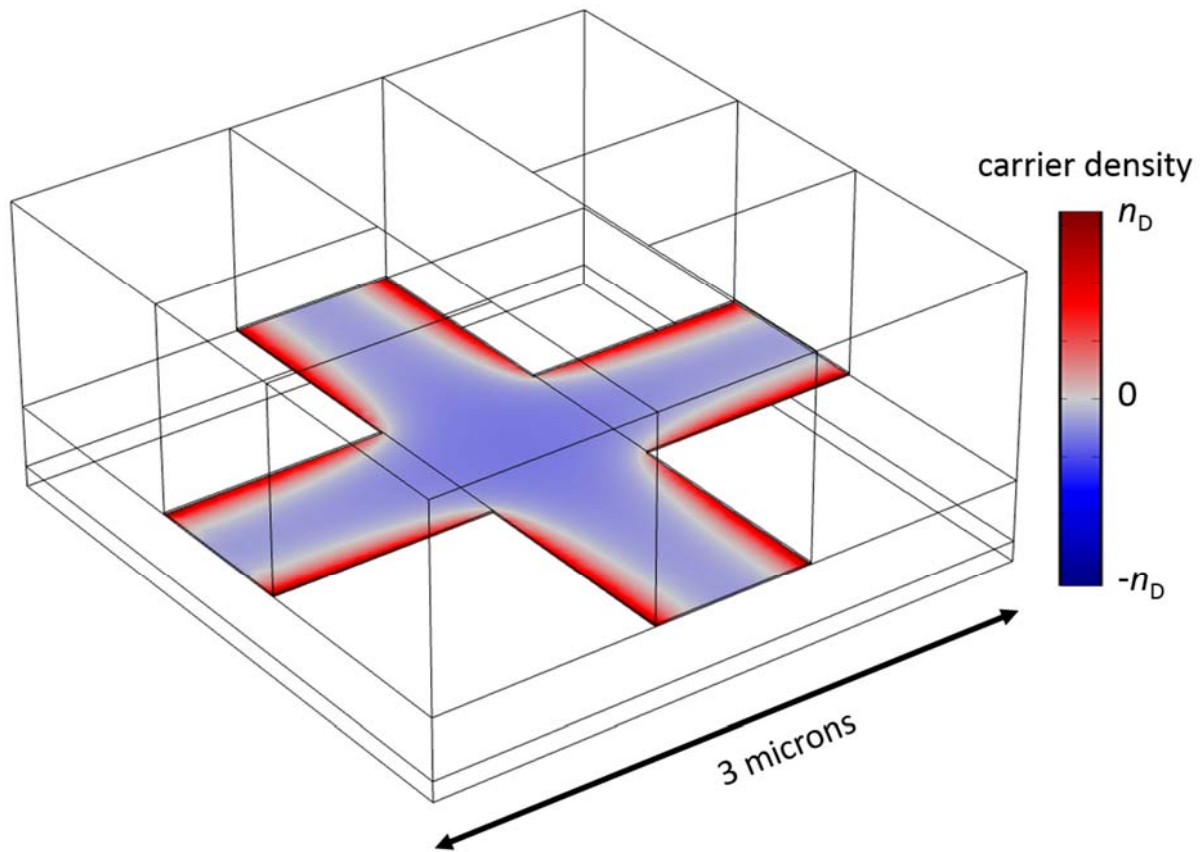

**Figure 1.** Calculated charge carrier density (colour scale) induced by a back gate in a cross-shaped graphene device with arms of width 800 nm over a 300 nm $SiO_2$ gate dielectric. The graphene is assumed to have an initial uniform doping $-n_D$ and the gate voltage is such that the gated graphene is charge neutral on average. The light grey regions represent the location of *p-n* junctions in the graphene. Periodic boundary conditions are used in the lateral directions.

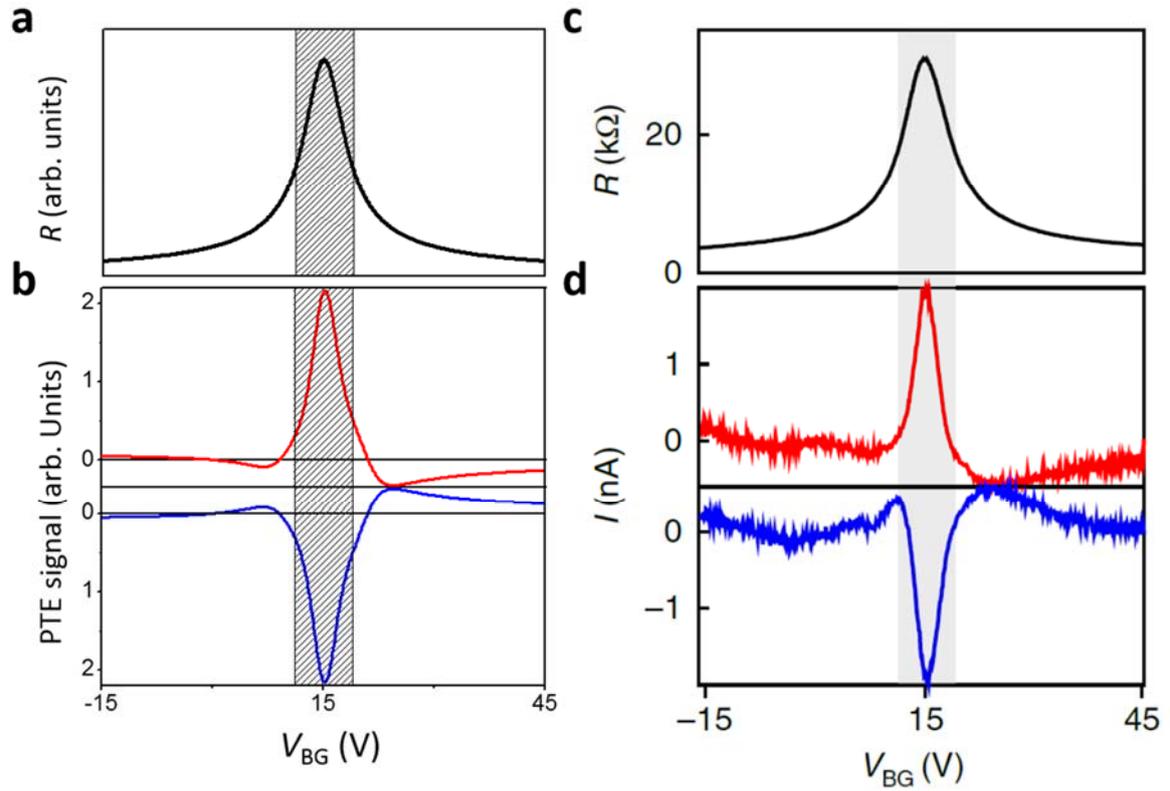

**Figure 2.** (a-b) Results of model discussed in text for resistance *R* and PTE signal as a function of gate voltage $V_{BG}$, respectively. (c-d) Resistance *R* and photocurrent *I* as a function of gate voltage, reproduced from Figs. 2e and 2f of [1].